\documentclass[12pt,preprint]{aastex}

\slugcomment{}

\shorttitle{Binary Protostars: Data}
\shortauthors{Connelley et al.}

\begin{document}


\title{The Evolution of the Multiplicity of Embedded Protostars I: Sample Properties and Binary Detections\footnote{The Infrared Telescope Facility is operated by the University of Hawaii under Cooperative Agreement no. NCC 5-538 with the National Aeronautics and Space Administration, Science Mission Directorate, Planetary Astronomy Program. The United Kingdom Infrared Telescope is operated by the Joint Astronomy Centre on behalf of the Science and Technology Facilities Council of the U.K. Based in part on data collected at Subaru Telescope, which is operated by the National Astronomical Observatory of Japan.}}


\author{Michael S. Connelley\altaffilmark{1}, Bo Reipurth\altaffilmark{2}, and Alan T. Tokunaga\altaffilmark{3}}


\altaffiltext{1}{Michael.S.Connelley@nasa.gov, NASA Ames Research Center, MS 245-6, Moffett Field, CA 94035}
\altaffiltext{2}{University of Hawai'i Institute for Astronomy, 640 N. Aohoku Pl., Hilo HI 96720}
\altaffiltext{3}{University of Hawai'i Institute for Astronomy, 2680 Woodlawn Dr., Honolulu, HI 96822}


\begin{abstract}

  We present the observational results of a near-infrared survey of a large sample of Class I protostars designed to determine the Class I binary separation distribution from $\sim100$~AU to $\sim5000$~AU.  We have selected targets from a new sample of 267 nearby candidate Class I objects.  This sample is well understood, consists of mostly Class I young stellar objects (YSOs) within 1~kpc, has targets selected from the whole sky, and is not biased by previous studies of star formation.  We have observed 189 Class I YSOs north of $\delta=-40^{\circ}$ at H, K and L$'$-bands, with a median angular resolution of 0\farcs33 at L$'$.  We determine our detection limit for close binary companions by observing artificial binaries.  We choose a contrast limit and an outer detection limit to minimize contamination and to ensure that a candidate companion is gravitationally bound.  Our survey uses observations at L$'$ rather than K-band for the detection of binary companions since there is less scattered light and better seeing at L$'$.  This paper presents the positions of our targets, the near-IR photometry of sources detected in our fields at L$'$, as well as the observed properties of the 89 detected companions (73 of which are newly discovered).  Although we have chosen contrast and separation limits to minimize contamination, we expect that there are $\sim$6 stars identified as binary companions that are due to contamination. Finder charts at L$'$ for each field are shown to facilitate future studies of these objects.

\end{abstract}

\keywords{binaries: general, stars: formation, stars: statistics, infrared: stars}



\section{Introduction}
  Ever since it was demonstrated that there must be physically bound pairs of stars and star clusters \citep{Mit1767}, the question of binary star formation has been an unsolved problem in astronomy.  \citet{Duq1991} reported that the solar-type main sequence binary frequency\footnote{The binary frequency is the total number of companion stars divided by the number of systems.} is $50\%\pm5.5\%$ for stars with periods from less than a day to over 10 million years without completeness correction, and 61\% after completeness correction.  Based on the statistics of the main-sequence binary population, \citet{Lar2001} concluded that ``stars seldom if ever form in isolation''.

   The binary frequency of T Tauri stars has also been carefully studied because they are young, there are a large number of them, and they are optically visible.   \citet{Rei1993} conducted an optical survey of 238 southern pre-main sequence stars and found a binary frequency of $16\%\pm3\%$ over the range of projected separations from 150 to 1800~AU.  More recently, \citet{Mat2000a} and \citet{Pat2002} tabulated the results of multiplicity surveys among pre-main sequence stars.  Overall, T Tauri stars are found to have roughly twice the binary frequency compared to main sequence solar type stars over the separation ranges covered by these studies.

\citet{Duc2004} and \citet{Hai2004} have published results from searches for embedded binary YSOs in nearby star forming regions.  \citet{Duc2004} found a binary frequency of $\sim26\%\pm8\%$ in the separation range from 110~AU to 1400~AU in a survey of 63 flat spectrum and Class I YSOs in the Taurus and Ophiuchus clouds.  \citet{Hai2004} observed a similar sample of 76 YSOs in the Perseus, Taurus, Chamaeleon, Ophiuchus, and Serpens clouds, finding a binary frequency of $18\%\pm4\%$ in the separation range from 300~AU to 2000~AU.  Combining both results, \citet{Duc2006} found a total of 19 companions to 119 stars, yielding a binary frequency of $16\%\pm4\%$ from 300 to 1400~AU.  This is roughly twice the binary frequency of main sequence stars over the same separation range, and is consistent with the binary frequency of T Tauri stars. 

   We have performed a major study of the binarity of Class I sources, which we present in this and a companion paper (Connelley et al. 2008, hereafter Paper II).  New observations were required to investigate a larger sample spread over a wider range of star forming regions at higher angular resolution than previous studies.  The goal of this paper is to present the sample of Class I YSOs we observed (section 2) and our observational results (section 3).  We discuss how we identified binaries and minimized contamination through a choice of contrast and separation limits (section 4).  We also include the properties of the binary companions that we found, including binary systems with strong color differences (section 5) that are analogues to infrared companions to T Tauri stars.  In Paper II we present the Class I binary separation distribution using the data presented here.  That paper also includes comparisons of the Class I binary separation distribution with the results of previous studies of Class I YSOs and other pre-main sequence stars, the evolution of the binary separation distribution within the Class I phase, and the dependence of the Class I binary frequency on the star forming environment.

\section{Sample Properties}
   We used a new sample of nearby mostly Class I YSOs described in \citet{Con2007}.  Briefly summarized, the sample was selected based on IRAS colors, coincidence with nearby dark clouds, and coincidence with a red (H-K$\gtrsim$1) 2MASS source.  Distance estimates, usually to the cloud hosting the protostar, were taken from the literature and are listed in Table 1, along with the source of the distance estimate.  The distance distribution (Figure 1) shows that most objects are within 1~kpc, with a median distance of 470~pc.  We do not have distance estimates to all of the targets in our sample, however several targets without distance estimates appear to be associated with well known clouds.  The spectral index distribution (Figure 2) shows that the majority of our targets have a spectral index \citep{Lad1991} greater than 0, and thus the majority are Class I objects.  However, there are a few known T Tauri stars in our sample.  For example, FS Tau A was observed since it is a companion to FS Tau B, which is deeply embedded. The sample's spectral index distribution has a median value of +0.79 and a mean value of +0.91.  These spectral indices were derived using only IRAS fluxes from the Faint Source Catalog if the target is included in that catalog, or the Point Source Catalog if not.  We used flux measurements from 12~$\mu$m to 100~$\mu$m, unless the 12~$\mu$m measurement is an upper limit, in which case the spectral index was calculated from 25~$\mu$m to 100~$\mu$m. Several IRAS sources have more than one near-IR counterpart in the IRAS beam, thus higher angular resolution far-IR observations may yield different values for the spectral index. The bolometric luminosities of all sources were calculated as described in \citet{Con2007}.  The dearth of sources with L$_{bol}>100$~L$_{\odot}$ suggests that relatively few of the stars in the sample are high mass stars.  This is expected since \citet{Con2007} selected against sources associated with HII regions.  Similarly, the dearth of sources with L$_{bol}<0.5$~L$_{\odot}$ shows that it is unlikely that there are many proto-brown dwarfs in the sample.  

   A goal of the sample selection process was to choose sources across the entire sky, without bias towards well known star forming regions.  Figure 3 shows the distribution of targets in Galactic coordinates.  The shaded area on the right is the part of the sky that is south of $\delta=-40^{\circ}$.  Targets in this region do not rise above 2 airmasses from Mauna Kea and, aside from two exceptions, were not observed.  Our targets are spread across all Galactic longitudes, with clumping in the Taurus/Auriga and Orion star forming regions.  These two regions are both below the Galactic plane and near the Galactic anti-center.  Figure 4 shows the arrangement of targets as seen from above the Galactic plane.  The targets that are in our sample, including those too far south for us to observe, are listed in Table 1.  

\section{Observations}
 \subsection{Target Selection}
    Not all of the targets in the sample were observed in the course of our study. Some of the objects in the sample are T Tauri stars, while others are Class 0 objects, or a filament or knot in a cloud.  A few of the stars in our sample are examples of what have become known as ``transitional'' objects, i.e. objects between Class I and T Tauri stars with a spectral index near 0.  These sources were observed as they satisfied our selection criteria and since the studies by \citet{Hai2004} and \citet{Duc2004} include such objects.  We did not attempt to observe the Class 0 objects since there is typically no flux in the near-IR. In some cases, MSX \citep{Pri2001} observations showed that the IRAS point source was a knot or a filament in a cloud, and not a true point source.  Such sources have no near-IR counterpart and were not observed.  

 \subsection{Observation Methods}
   Previous studies of the Class I binary frequency (Haisch et al. 2004 and Duch{\^e}ne et al. 2004) searched for binary companions at K-band.  We found that the seeing was better and more stable at L$'$ than at K-band, and that reflection nebulae (which tend to have blue colors) are much less of a problem at L$'$.  The bright sky background at L$'$ also made it more difficult to see stars without an IR excess, reducing the effect of background star contamination.  We therefore focused our search for binary companions on our L$'$ observations and used the K-band and H-band observations for additional photometry.  

   Since we wanted to observe a large number of targets from H through L$'$, we chose telescope/instrument combinations that have this capability in one instrument, have good image quality, and have a suitable plate scale.   We used the UH~2.2~m telescope with QUIRC (1024$^{2}$ HgCdTe 1-2.5~$\mu$m 3\arcmin ~FOV, Hodapp et al. 1996), the NASA IRTF with the SpeX guider (512$^{2}$ InSb 1-5~$\mu$m 1\arcmin~FOV, Rayner et al. 2003) and NSFCam2 (2048$^{2}$ Hawaii-2RG 1-5.5~$\mu$m 82\arcsec ~FOV), UKIRT with UIST (1024$^{2}$ InSb 1-5~$\mu$m 1\arcmin~FOV, Ramsay Howat et al. 2004), and Subaru Telescope with CIAO (1024$^{2}$ InSb 1-5~$\mu$m 22\arcsec~FOV, Murakawa et al. 2004) and IRCS (1024$^{2}$ InSb 1-5~$\mu$m 1\arcmin~FOV, Tokunaga et al. 1998 and Kobayashi et al. 2000).  Table 2 lists which telescopes were used on which nights.   The majority of our observations used UKIRT and UIST, primarily due to the availability of observing time.  Due to UKIRT's north declination limit of +60$^{\circ}$, IRTF observations targeted sources north of this limit up to the north declination limit of IRTF (+70$^{\circ}$) .  We used Subaru to observe targets north of this, targets for which we did not get good image quality with IRTF, and to observe targets with adaptive optics (AO).

     Dithering was used for all observations in order to remove bad pixels and the detector flat field effects.  A 3$\times$3 dither pattern was typically used, the size of which was usually 5\arcsec~ to 10\arcsec~ and depended on the field of view of the instrument and the availability of guide stars.  In the case of L$'$ observations, coadds were used to increase the effective integration time per dither position to $\sim$20~s to increase observing efficiency.  Standard stars that have been observed by UKIRT through the MKO filter set (Simons \& Tokunaga 2002, Tokunaga \& Simons 2002) were selected from the UKIRT faint standard star list, and were observed for photometric calibration.  Furthermore, the instruments we used have MKO filters, and thus all of our observations are in the MKO photometric system.

   All data were reduced using the following procedure except in the case of the IRTF data, where the data were first divided by the product of the number of coadds and the number of non-destructive reads.  A dark frame was made by averaging together 10 individual darks of the same exposure time as the science data.  This dark was then subtracted from each target frame.  To make the sky frame, each dark subtracted frame was scaled to have the same median value, then averaged together using a min-max rejection.  The resulting sky frame was then normalized using the median value of the pixel counts.  Each dark subtracted (non-scaled) target frame was divided by this normalized sky flat.  The median sky value for each frame was subtracted from each frame to set the average background counts in each frame to 0.  The images were then aligned and averaged together using an average sigma clipping rejection.  In addition, images with better than average resolution were combined into a higher resolution image.  This rejects images where the seeing was particularly poor or where there was a guiding error.  Since most of the L$'$ ``sky'' brightness is from the telescope, the procedure we used was not optimal for making a true L$'$ flat field.  However, the L$'$ flats we used were effective for removing the detector's flat field response.

   For this project, having the best angular resolution possible was critical.  Particular attention was paid on maintaining the best focus possible.  In the case of our IRTF observations with the SpeX guider, the image resolution was often limited by aberrations either in the telescope or in the instrument.  Aberrations in UIST on UKIRT also occasionally limited our resolution at K, but rarely at L$'$.  We used the 0\farcs06 plate scale in UIST in order to be able to use a longer exposure time at L$'$ and to better sample the PSF.  The resulting 1\arcmin~FOV also allowed objects within 4500~AU of the target to be in the field of view for the closest targets.  The angular resolution distributions at H, K, and L$'$ are presented in Figure 5.  The median FWHM was 0\farcs609 at H-band, 0\farcs543 at K-band, and 0\farcs335 at L$'$-band.  The L$'$-band median FWHM includes our AO observations. 

 \subsection{AO Observations}
   The selection of targets in nearby dark clouds naturally selected against nearby bright stars that could be used as AO guide stars.  To find sources with a suitable visual guide star, we searched through the USNO-B1.0 catalog \citep{Mon2003} for stars within 40\arcsec~ of the near-IR source that are brighter than R or I=16.  The objects that we observed with AO are presented in Table 3.  To reduce the chance of reflection nebulosity interfering with our search for very close companions, we only observed sources with no resolved nebulosity in our seeing-limited L$'$-band data.  

   There are a number of cases where enough of the visible light from the YSO is able to escape the cloud to use the YSO itself as a guide star.  This raises the possibility that the AO observed sub-sample is, on average, older and more evolved than the sample as a whole.  We used the Kolmogorov-Smirnov test to determine if the AO observed sub-sample is different from the whole sample based on the spectral index and bolometric luminosity distributions.  The whole sample and the AO observed sub-sample are not statistically different with regards to spectral index or bolometric luminosity at the 3~$\sigma$ level.   

 \subsection{Target Fields}
    Figure 6 shows a 20\arcsec$\times$20\arcsec~ (3000~AU to 10$^{5}$~AU, depending on distance) field around each target at L$'$\footnote{This figure could not be included in the astro-ph upload due to their file size limit.  This paper with all figures can be downloaded from http://homepage.mac.com/mconnelley/FileSharing1.html.  Thank you.}.  For targets where multiple near-IR sources do not fit within this field, more than one field is shown.  Each near-IR source is labeled with a number that corresponds to that object's photometry presented in Table 4 if there is more than one object in the field.  Inset images show regions of interest in more detail.  In some cases, the primary star has been subtracted to better show a companion star in the inset image.

\subsection{Photometry}
   We obtained H, K, and L$'$-band data in order to use the near-IR colors to separate embedded YSOs from foreground or background stars.  We used archival CFHT Skyprobe data to ensure that the data we used for photometry were taken under photometric conditions, characterized by a stable attenuation measurement near 0 throughout the night. On nights that were non-photometric, the photometry was calibrated using field stars in our photometric data or in 2MASS.  If we used 2MASS for H and K-band photometry and we have our own L$'$ observations, the variability of our targets affects the accuracy of the colors that we derive, since the target may have varied in brightness between the 2MASS observations or between the 2MASS and our observations.  We also converted the 2MASS photometry to the MKO system.  Aperture photometry was performed using IMEXAMINE in IRAF using five aperture sizes (typically 0\farcs9, 1\farcs2, 1\farcs5, 1\farcs8, 2\farcs1) while maintaining the same buffer and sky annulus width (both typically 1\farcs8) for each aperture size.  The same procedure was used for our standard stars.  We compared the brightness of the target and the standard using the same aperture size to derive a magnitude estimate for each of the five aperture sizes.  We then averaged these five estimates together, taking the standard deviation of these measurements as the accuracy to which we could measure the photometry of that individual source given the quality of the data.  We made an airmass correction plot using our standard star data.  We used the standard deviation of the standard star photometry from the best fit linear airmass extinction curve as the lower limit to our photometric errors.  This error was combined with the individual measurement error via a Pythagorean sum (root of summed squares) to estimate the total photometric error for each object in each filter ($\delta$H, $\delta$K, and $\delta$L$'$ in Table 4).  We used the airmass extinction values in \citet{Kri1987} for our airmass correction.

\section{Binary Detection}

   All binary stars were found by visual inspection of our images.  We found that Fourier filtering of our PSF subtracted images (described below) did not enhance the visibility of close or faint companions since the PSF subtraction residuals had the same spatial size as a real companion.  We did not attempt to use an automated star finding program on account of our previous experience with programs such as DAOFIND.  As an example, if the search parameters were set to find faint stars, then it would also identify positions without a star.  Since our fields only had a few objects, a star finding program was not necessary.  

 \subsection{Contrast Limit}
   Visual surveys for binary stars are always sensitive to contamination from background stars.  One way to minimize background star contamination is by adopting a contrast limit, such that any star fainter than the limit is not considered as a potential companion since the possibility of such a faint star being background contamination is unacceptably high and the chance of it being a real companion is acceptably low.  \citet{Hai2004} used a contrast limit of $\Delta$K=4, whereas \citet{Rei1993} adopted a $\Delta$z=5 contrast limit.  \citet{Duq1991} found that nearly all main sequence binary stars with a solar-type primary have a mass ratio greater than 10:1.  In light of this, we should choose an L$'$ band contrast limit that allows for all binaries with a mass ratio greater than 10:1.  \citet{Rei1993} state that for coeval stars on the Hayashi track, the flux ratio approaches the mass ratio as the wavelength increases, with these ratios being effectively equal at 2.2~$\mu$m.  Consider a binary system with a mass ratio (and thus a photospheric flux ratio) of 10:1, where only the primary star has an infrared excess.  In this case, the primary star's infrared excess can be up to three times greater than its photospheric flux without the observed flux ratio of the binary exceeding 40:1.  Thus, a contrast limit of $\Delta$L$'$=4 satisfies our criteria for not excluding a significant number of real companions.
   
 \subsection{Artificial Binary Detection and the Inner Detection Limit}
    The angular resolution of the images, the contrast between the primary star and the companion, and to a lesser degree the plate scale of the camera affected how close we were able to detect a companion star.  PSF fitting and subtraction was done with our L$'$ data only to reveal very close and faint companions.  The most successful PSF model was another field star in the same image.  Since the image of the field star and target star were taken simultaneously, the PSFs of the two are nearly identical, and thus the field star is an excellent PSF model.  However, this method could only be used rarely since the probability of another bright star being in the field of view is quite low.  We usually used stars observed just before and just after the target to be subtracted, and combined these two PSFs into a model PSF for the one to be subtracted.  The typical peak counts of the residual after PSF subtraction using this method are about 4\% of the PSF's peak counts and are typically found about 1 FWHM from the center of the PSF.  There were cases where a star had excessive PSF residuals, either from a poor fit or due to scattered light off of circumstellar material at L$'$.  Scattered light was much less of a problem at L$'$ than at K but is still present, especially very close too the star.  Excessive PSF residuals affected how close we could detect fake binary companions (described below), and this is reflected in our inner detection limits.

   Knowing when companion stars could have been missed is nearly as important as detecting the companions themselves.  Our data are less sensitive to close and faint companion stars.  Thus, for each target, we needed to determine the closest separation that a companion of a given contrast could have been found so that we could later correct for our incomplete sensitivity to close and faint companions.  To do this, we inserted artificial companions at a range of contrasts ($\Delta$L$' =$~1, 2, 3, and 4 magnitudes fainter than the primary star) into the PSF subtracted image of each target star, regardless of whether it has a real binary companion.  At each contrast level, we inserted 20 artificial companion stars, one at a time, at a known radius but at a random position angle into the PSF subtracted image.  The image of each artificial binary was viewed for 1~s to ensure that we could easily and confidently find the artificial companion.  The artificial companion also had to be easy enough to recover so that, if we were examining real data, we would confidently believe that we had found a companion star.  Each artificial binary image was followed by an image of blank sky, also for 1~s, because we found that it was too easy to see the artificial companion ``jump'' around the image if images of artificial binary companions in different locations were viewed consecutively.  If the companion could be recovered at least 19 out of 20 times, then the artificial companion would be inserted at a closer separation and the test repeated until the artificial companion could not be reliably recovered 95\% of the time.  The inner detection limit is determined to be the closest separation where the artificial companion could be reliably detected at least 95\% of time.  This test was repeated for each of the four contrast levels mentioned above, and for each individual star.  
   
     This method has the disadvantage that we know at what separation to expect artificial companions to be found.  However, if we placed artificial companion stars at random separations and random position angles, most of the artificial stars would be inserted at a separation either too close to be recovered, or far enough away to be trivially detected.   Even in the case where the artificial star is inserted at a random separation, we are most interested in artificial companions in the separation range where it is possible but difficult to detect the artificial companion.  The method we used has the advantages that it quickly identifies the inner separation limit, and it uses the same method used for identifying real binary companions.  Table 5 lists the binary systems that we identified. Table 6 lists the inner detection limit for each star at four contrast levels, as well as the outer companion acceptance limit (described below) at each of the four contrast levels.

 \subsection{Outer Detection Limit}
    The purpose of imposing an outer separation limit, beyond which no object would be considered as a companion, is to ensure that all candidate companions are likely to be gravitationally bound companions to the primary star and to help eliminate background star contamination.  Duch{\^e}ne et al. (2004)  used an outer limit of 10\arcsec~ (1400~AU~at the distance of their targets).  They argue that this outer limit is much smaller than the typical size of a typical core in the regions they observed, thus these binaries are likely to have formed from the collapse of the same core or filament.  \citet{Rei1993} used an outer limit of 1800~AU.  They argue that the typical star-to-star separation in a low density star forming region is $\sim$20,000~AU, and is $\sim$10,000~AU in a high density region such as the Trapezium cluster.  As such, 1800~AU is an order of magnitude smaller than the typical star-to-star separations for the regions that the targets observed by \citet{Rei1993} are in, and thus they argue that these companions are likely to be gravitationally bound.   

    There are a handful of well known common proper motion binary stars with very wide separations that are believed to be gravitationally bound.  Perhaps the first star to be recognized as a real binary (versus an optical double) is $\beta$ Capricorni \citep{Mit1767}, which has a projected separation of 9400~AU.  $\epsilon$ Lyrae 1 and 2 have a common proper motion and a projected separation of 13,000~AU \citep{Bur1978}.  While it is rare for a companion to have a separation in excess of 2000~AU, it is possible for such widely separated stars to be gravitationally bound.  Furthermore, the mean velocity dispersion of CO gas in the Taurus clouds is 1.4 kms$^{-1}$, and the observed radial velocity dispersion of Class I protostars is consistent with this value \citep{Cov2006}.  At this velocity, it would take $1.7\times10^{4}$ years, or roughly the Class 0 life time, to drift 5000~AU.  Thus, close but gravitationally unbound stars should be more than 5000~AU apart by the time they are visible in the near-IR as Class I YSOs.  We accept companions with projected separation as great as 5000~AU in order to include widely separated companions with confidence that they are likely to be gravitationally bound.

   The probability of background star contamination within a projected separation of 5000~AU could exceed 5\%, which we consider unacceptably high.  This is particularly true in regions near the Galactic center.  We used star counts in our L$'$ data to estimate the probability of contamination for each target.  We counted all stars in our L$'$ images with near-IR colors consistent with field stars.  Since there are many fields with no apparent field stars, we grouped these fields into seven regions of Galactic longitude and latitude to improve the count statistics.  Having also derived the L$'$ apparent magnitude distribution for all stars in all fields, we used the star counts in a given region to estimate the density of field stars less than L$'=4$ magnitudes fainter than each of the primary stars in that region.  We used this density and Equation 1 from \citet{Cor2006} to estimate the radius from the star where the probability of contamination exceeds 5\%.  This angular radius is:

\begin{equation}
  \theta = \sqrt{-ln(1-P)/\pi\Sigma}
\end{equation}

  where $\theta$ is the angular radius with the probability of contamination P (in our case, P=0.05), and $\Sigma$ is the surface density of field stars in that region of Galactic longitude and latitude that are less than L$'$=4 magnitudes fainter than the YSO in question.  If the 5\% contamination radius has a projected radius less than 5000~AU, then the contamination limited radius was used as the outer limit for accepting companions.  Otherwise, 5000~AU was used.  Although the maximum chance of contamination is 5\%, the average chance of contamination within the adopted outer separation limit is 3.0\%.  Thus we expect there are $\sim$6 (189$\times$0.03) stars identified as binary companions that are background contamination stars.  We note that there are a number of fields where the stellar density is so high that we can not confidently identify which object in the field, if any, is the near-IR counterpart to the IRAS source.  These targets were thrown out and not considered as having been observed.
  
 \subsection{Color Selection Criteria}
      The goal of observing our targets in three bands was to use the location of each star observed (target, candidate companion, or background star) on an H-K vs. K-L$'$ color-color diagram to minimize the chance of background star contamination.  The H-K vs. K-L color-color diagram is divided into three main regions: a region of forbidden colors to the left of the reddening vector from the main sequence, a region of colors consistent with a reddened T Tauri star, and a region of colors consistent with a protostar having an IR excess greater than a T Tauri star (the region for reddened main sequence stars is very narrow).  The colors of unreddened T Tauri stars (in the CIT photometric system, not in the MKO system) were adopted from \citet{Mey1997}.  The direction of the reddening vector was derived from interstellar extinction values (assuming R=5, characteristic of dense clouds) taken from \cite{Mat2000b}.  Using these values, the H-K reddening is 0.079 per magnitude of A(V) extinction, and the K-L (not MKO L$'$) reddening is 0.066 per magnitude of A(V) extinction.  Thus, the reddening vector has a slope of 1.20 on the H-K vs. K-L color-color diagram.  

   We excluded those stars whose colors are consistent with a reddened or unreddened main sequence star or with forbidden colors, with due caution.  The colors of a close companion star are difficult to determine accurately due to the proximity of the brighter primary star.  Photometric errors and variability affect the measured colors.  Reflection nebulosity can strongly affect the observed colors of a star, especially at H and K bands. Nebulosity makes the star appear bluer, and may not be spatially resolved.  The colors of a protostar can range from the forbidden region (if the near-IR flux is dominated by scattered light) to the region characteristic of objects with a strong IR excess.  As such, the color information had to be used with other selection criteria, such as the proximity to the IRAS position and the presence of a spatially resolved reflection nebula, to decide if a star is likely to be an embedded YSO or background contamination.  Star counts were used along with colors since colors alone are not sufficient to mitigate the chance of background star contamination. 

 \subsection{Discovery Space}
   Choosing which candidate companion stars would be retained for further consideration depended on several factors.  Stars with H-K and K-L$'$ colors near 0 are likely to be foreground stars and were excluded.   An optical or IR reflection nebula is a clear sign that the object in question is physically associated with the cloud.  Accurate colors often could not be determined for very close companions.  Given the very low probability of contamination at such close separations, these candidate companions were kept.

    Figure 7 shows the range of separations and contrasts over which we actually found binary companions.  The number of companions versus log(separation/1\arcsec) is relatively constant.  When plotted against linear separation (arcseconds), most of the binaries have separations less than 3\arcsec.  Also, for most of the range of separations, we are not less sensitive to fainter companions than brighter ones.  It is only within a few times the FWHM (typically less than 1\arcsec) that we are less sensitive to faint companions due to the glare of the primary star and, in a few cases, nebulosity.  At wider separations, we can be less sensitive to faint companions since our data are not always deep enough to detect companions $\Delta$L$'=4$ fainter than the primary star. 
 
\section{Binary Color Differences}
   A large number of our binary systems have very different colors between the two components, a situation analogous to the infrared companions of T Tauri stars.  The prototypical case, T Tauri N and S, differ in their H-K colors by 1.39 magnitudes and in their K-L colors by 2.0 magnitudes \citep{Ghe1991}.   \citet{Zin1992} estimated that roughly 10\% of T Tauri binary stars have an IR companion.

   To find Class I analogues to T Tauri IR companions, we considered the difference in H-K and K-L$'$ colors between the components of binary systems.  This examination is limited to objects for which we have photometric data and where the binary is sufficiently well resolved that we have accurate photometry on each component.  We were able to derive 49 H-K color differences and 59 K-L$'$ color differences.  The color difference distributions are shown in Figure 8.  Both color difference distributions are centered near a color difference of 0, with the median $\Delta$(H-K)=0.016, and the median $\Delta$(K-L$'$)=0.245.  The $\Delta$(K-L$'$) distribution is slightly wider than the distribution of $\Delta$(H-K), the standard deviations being 1.11 and 0.89, respectively. Thus, there is no statistically significant preference towards the primary star (defined as the brightest star at L$'$) or the companion being redder. We find that 6/59 ($10.2\%^{+5.6\%}_{-3.9\%}$) of our Class I binaries have a K-L$'$ color difference more extreme than the T Tauri system, and 9/56 ($16.1\%^{+6.4\%}_{-5.0\%}$) have a H-K color difference more extreme than the T Tauri system, including 7 targets where we have a lower limit on the H magnitude of one of the components.  We note that only scattered light was detected at H and/or K-band for several targets, which naturally affects the observed colors.  We find that protostellar analogues to T Tauri IR companions are quite rare.  These values are consistent with the fraction of T Tauri stars that have an IR companion, suggesting a similar origin.  

\section{Summary}
   We have presented the results of a near-IR survey for binary stars in a new sample of nearby Class I protostars.  The purpose of this paper is to make our observations available to the community, to stimulate follow-up research on these protostars, and to present data on protostellar binary stars for detailed statistical analysis that is presented in Paper II.  This survey is distinguished by its well determined sample properties, large sample size, and choice of using L$'$ observations to identify protostellar binary companions.  We found 89 companion stars to 189 primary stars, 78 of which are within a projected separation of 5000~AU and have a contrast less than $\Delta$L$'$=4 magnitudes.  We have empirically determined our companion detection limits to account for our incomplete sensitivity to binary companions.  Separation and contrast limits were chosen to minimize the chance of background star contamination.  The average chance of background star contamination is 3.0\%, and we expect there are 6 stars identified as binary companions that are contamination.  Near-IR colors were used to identify contaminant stars and we showed that infrared companions are as rare among Class I YSOs as they are among T Tauri stars.  
   
\acknowledgments

\emph{Acknowledgments}  We thank the referee for helpful comments.  This research has made use of the SIMBAD database, operated at CDS, Strasbourg, France.  This research has made use of NASA's Astrophysics Data System.  This publication makes use of data products from the Two Micron All Sky Survey, which is a joint project of the University of Massachusetts and the Infrared Processing and Analysis Center/California Institute of Technology, funded by the National Aeronautics and Space Administration and the National Science Foundation.  This research was supported by an appointment to the NASA Postdoctoral Program at the Ames Research Center, administered by the Oak Ridge Associated Universities through a contract with NASA.

\clearpage
\begin{figure}
\plottwo{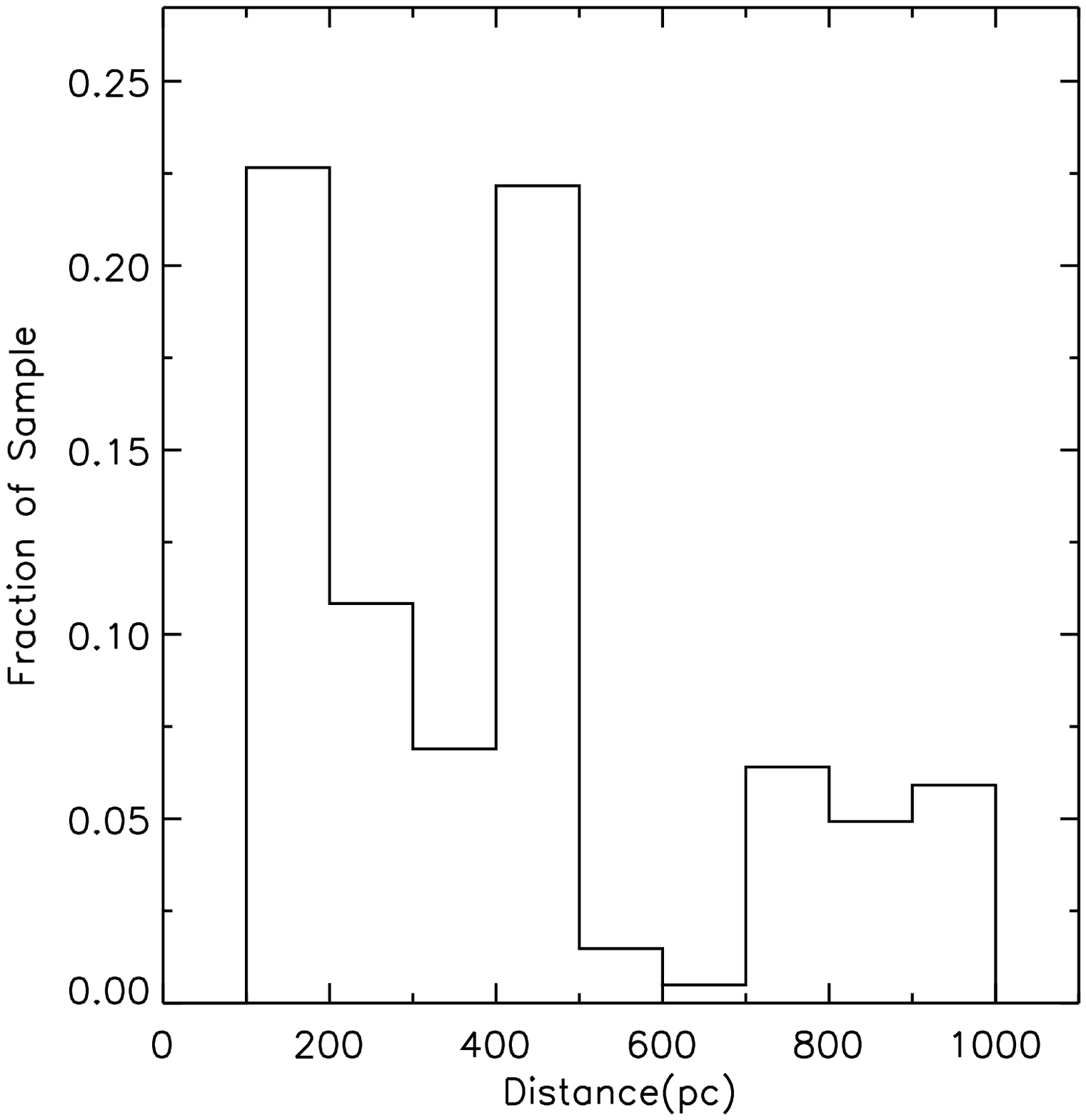}{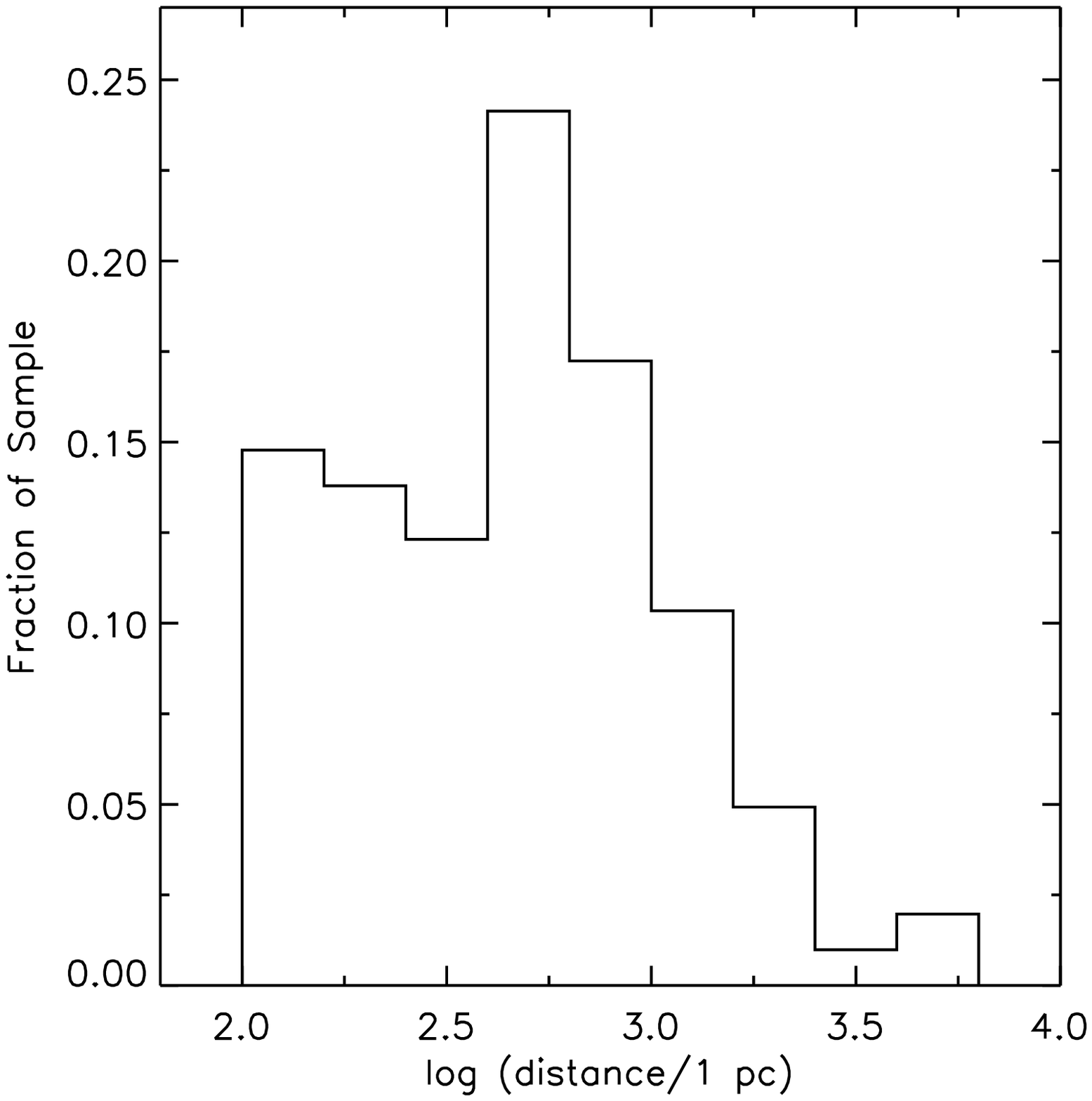}
\caption{Distance distribution for our sample.  The left panel shows the distance distribution for our sample on a linear scale out to a distance of 1~kpc.  The right panel presents the same data on a log scale, including targets as far as 6~kpc.  Our sample has a median distance of 470~pc and most objects are within 1~kpc.\label{this}}
\end{figure}

\clearpage
\begin{figure}
\plotone{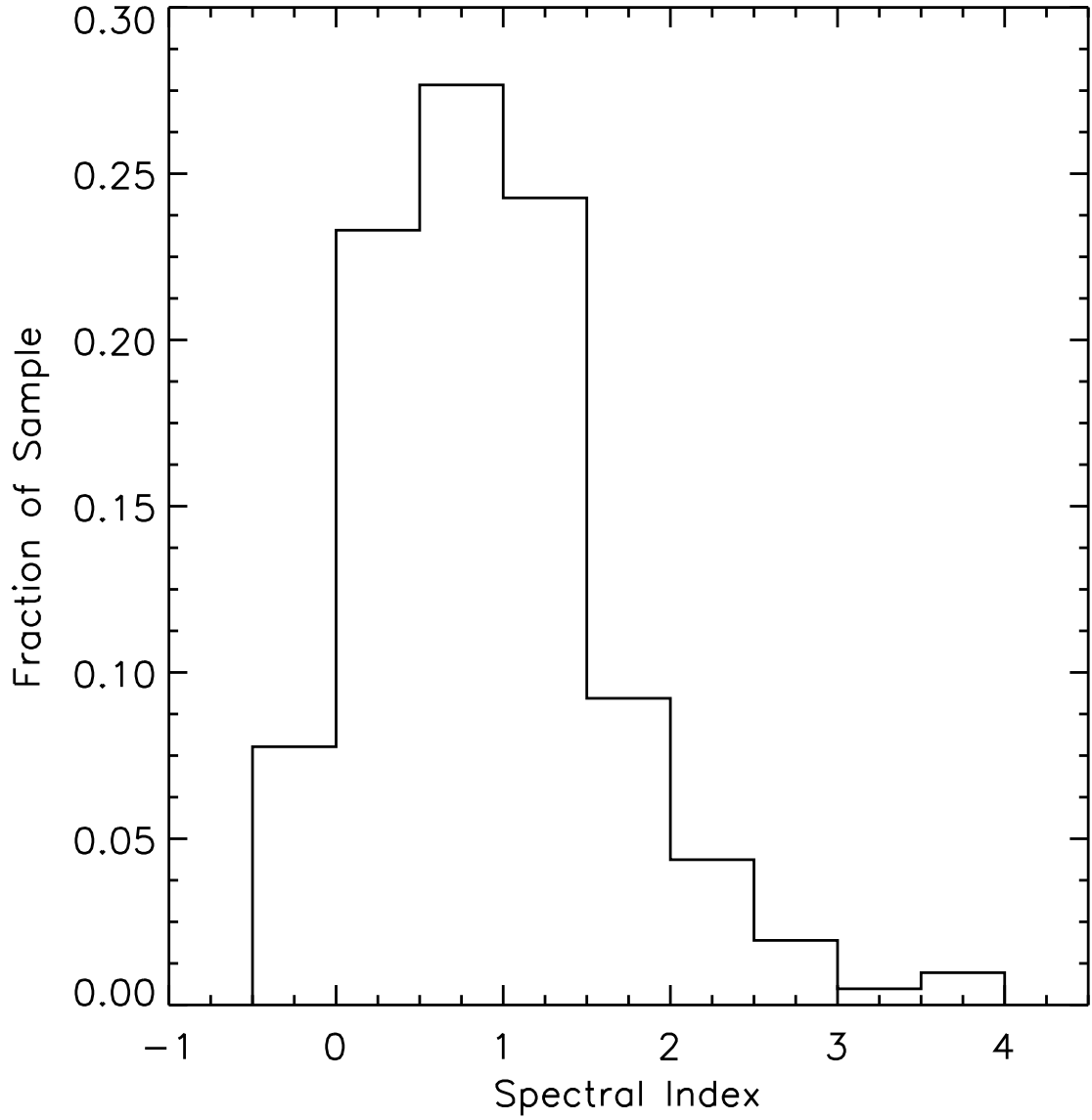}
\caption{Spectral index distribution. IRAS 12~$\mu$m to 100~$\mu$m fluxes were used to calculate the spectral index, using the method described by \citet{Lad1991}.  Our sample has a median spectral index distribution of +0.79, thus nearly all of our sources are Class I YSOs. \label{that}}
\end{figure}


\clearpage
\begin{figure}
\plotone{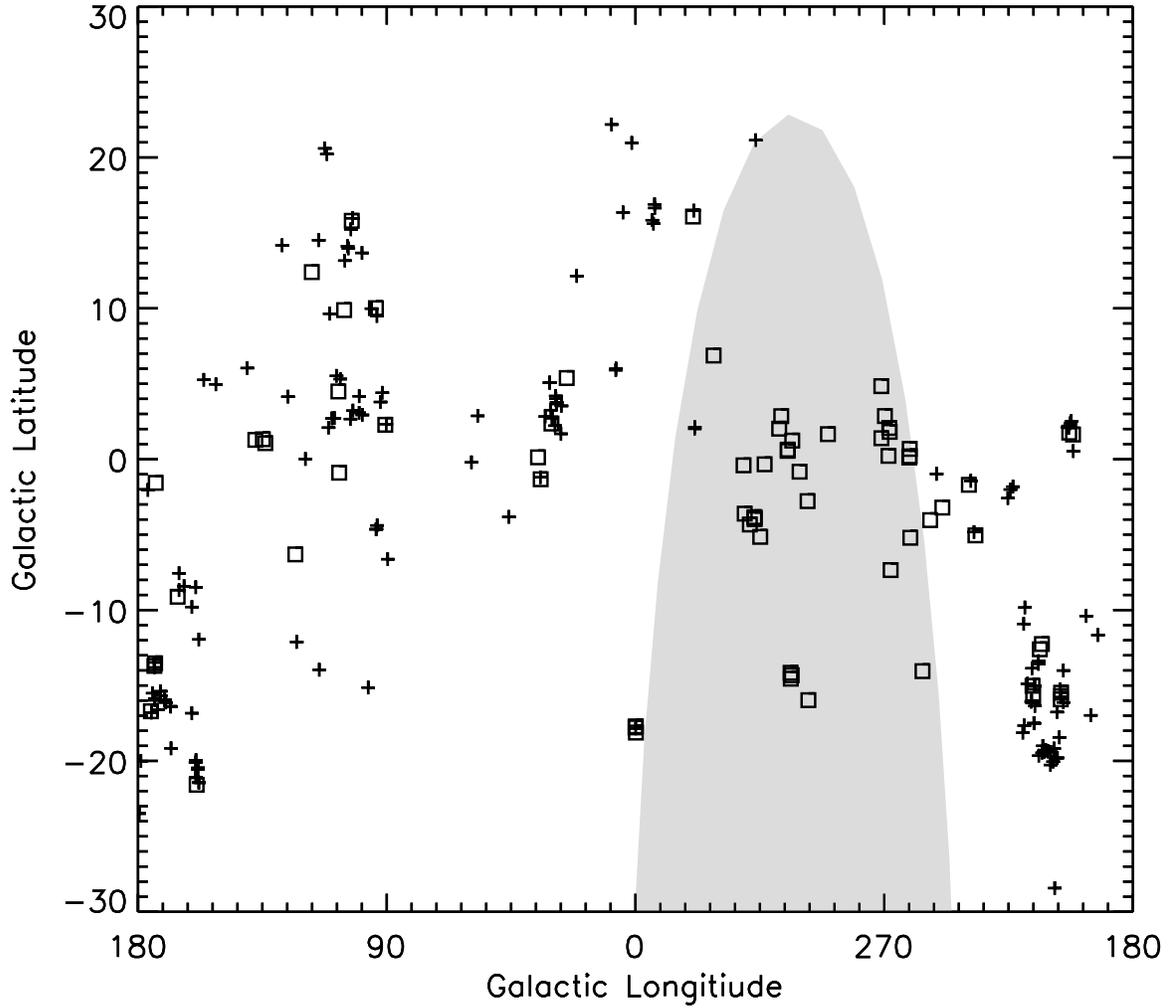}
\caption{Location of our Class I sources in Galactic coordinates.  The crosses are the targets we observed and the squares are targets that we did not observe, usually because they are too far south, there is no embedded near-IR counterpart to the IRAS source, or the source is an embedded cluster.  The shaded area to the right is south of $\delta=-40^{\circ}$, and never rises above 2 airmasses from Mauna Kea.  All of our targets are within $30^{\circ}$ of the Galactic plane. \label{fig2.3}}
\end{figure}

\clearpage
\begin{figure}
\plottwo{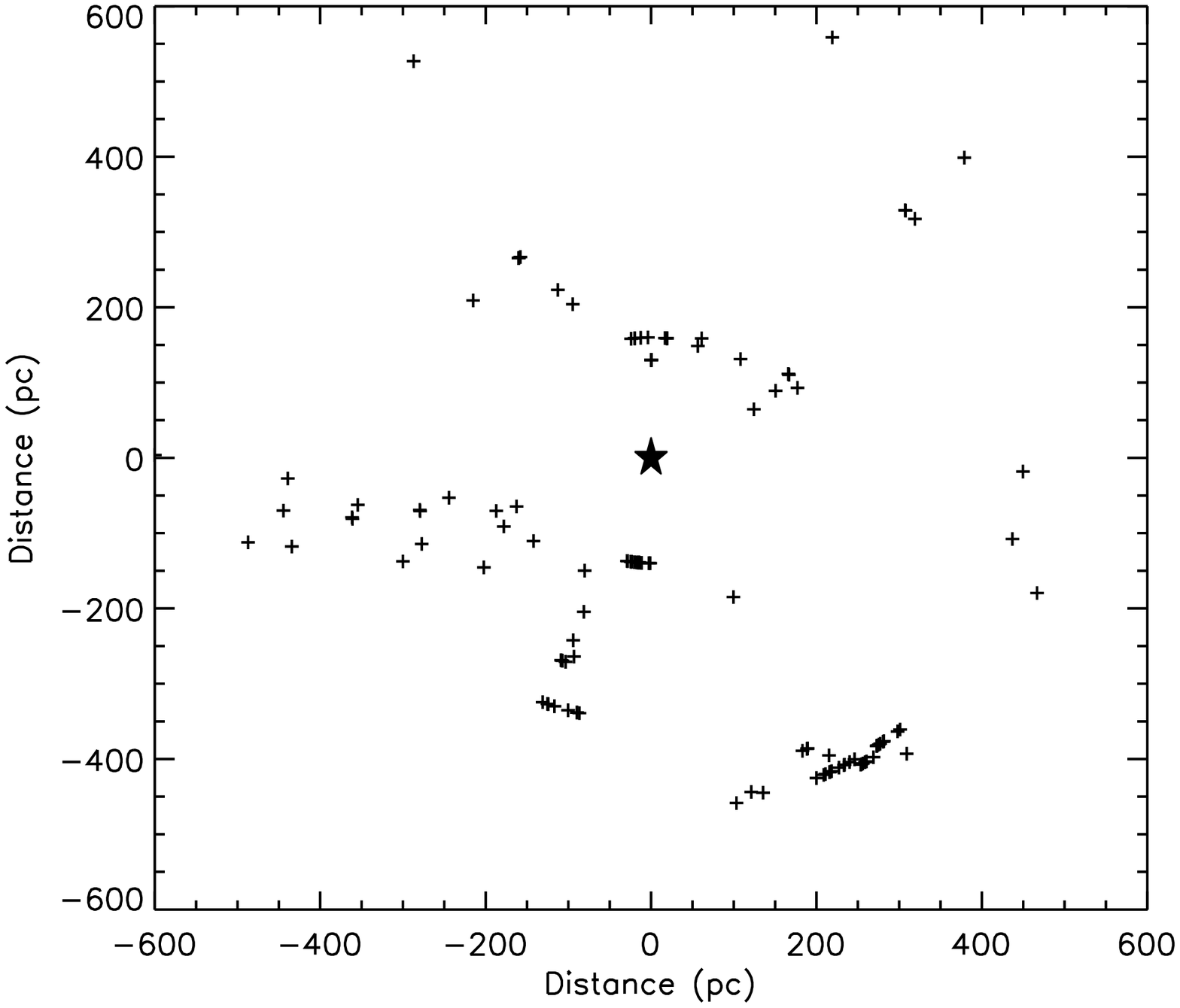}{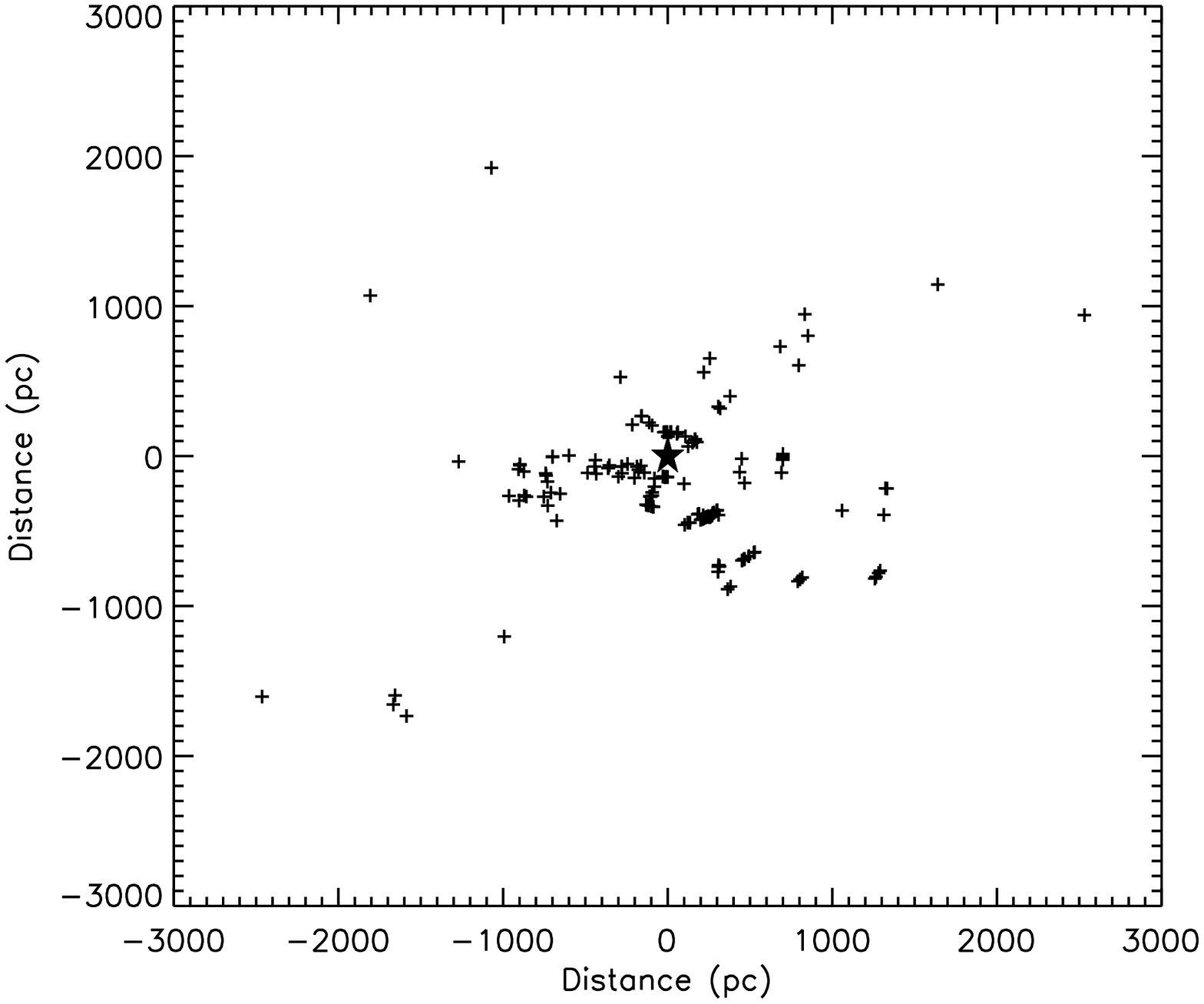}
\caption{Location of our Class I sources looking down on the Galactic plane.  The left panel shows sources out to a radius of 600~pc while the right panel shows sources out to a radius of 3~kpc.  The Sun is represented by the star symbol in the center of the figure.  The Galactic center is towards the top, the Taurus star forming region is just below the Sun, and the Orion star forming region is to the lower right. \label{fig2.3}}
\end{figure}

\clearpage
\begin{figure}
\plotone{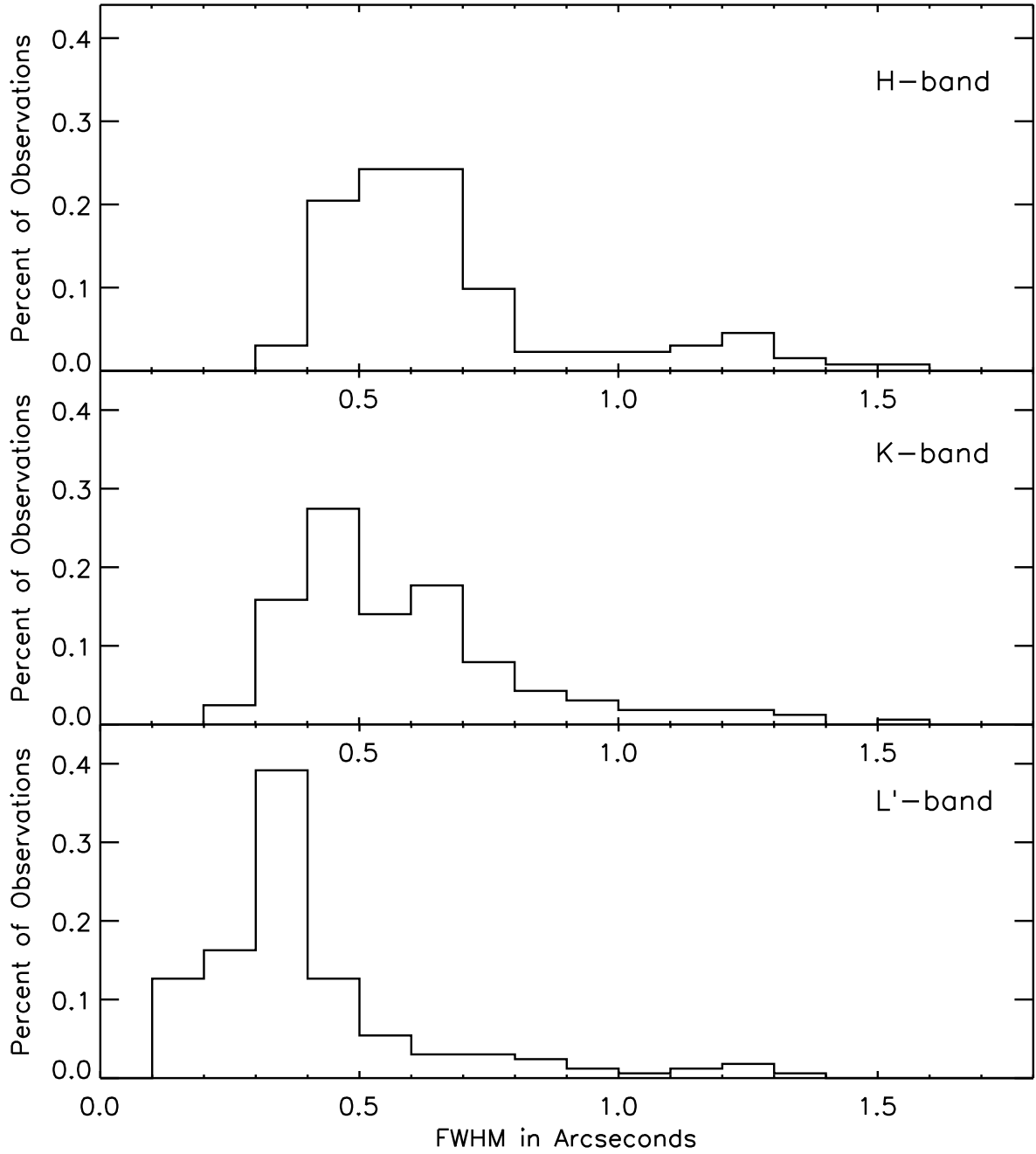}
\caption{Angular resolution distributions at H, K, and L$'$. The median angular resolution (FWHM) is 0\farcs609 at H, 0\farcs543 at K, and 0\farcs335 at L$'$. \label{fig2.1}}
\end{figure}



\clearpage
\begin{figure}
\plottwo{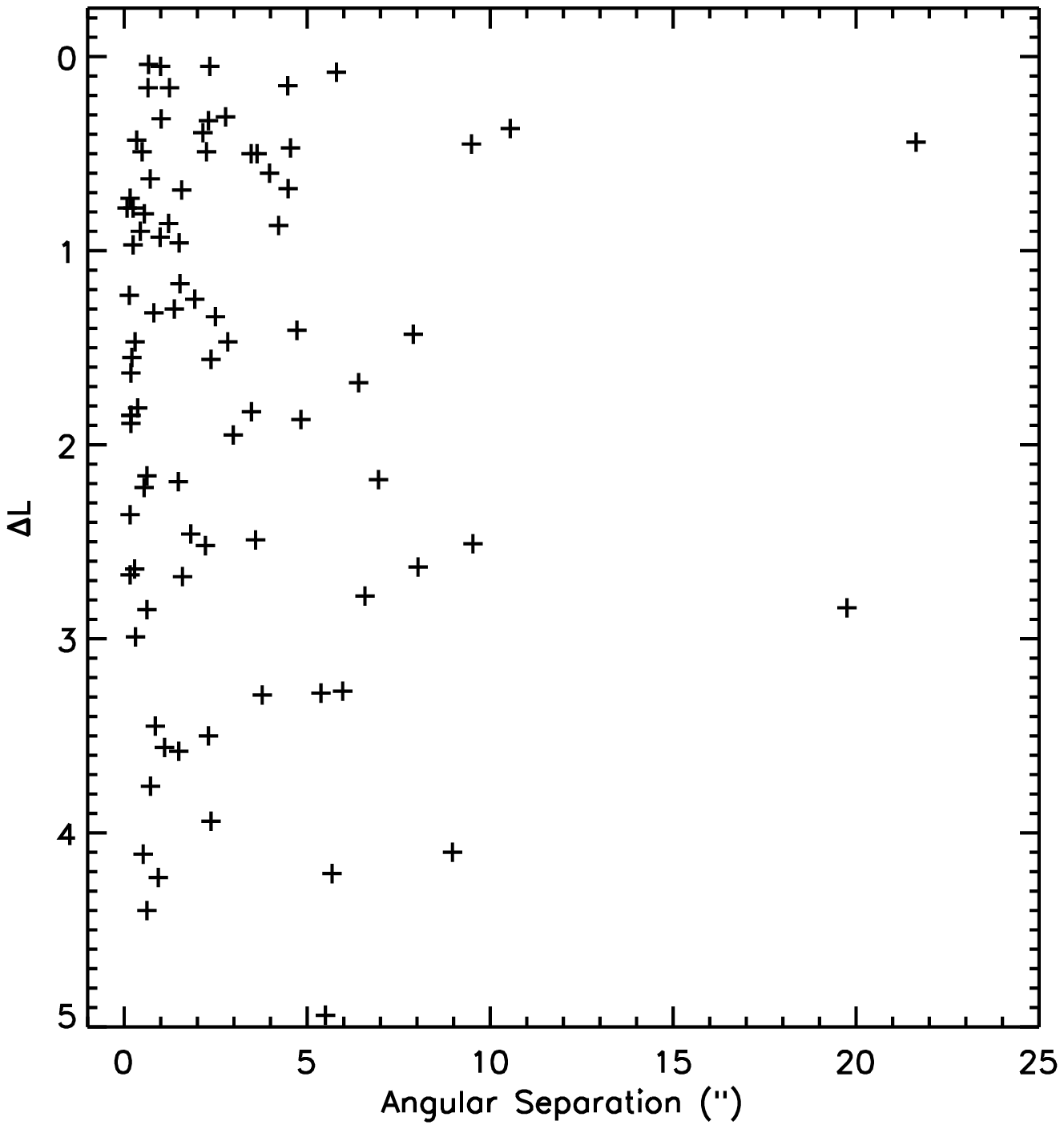}{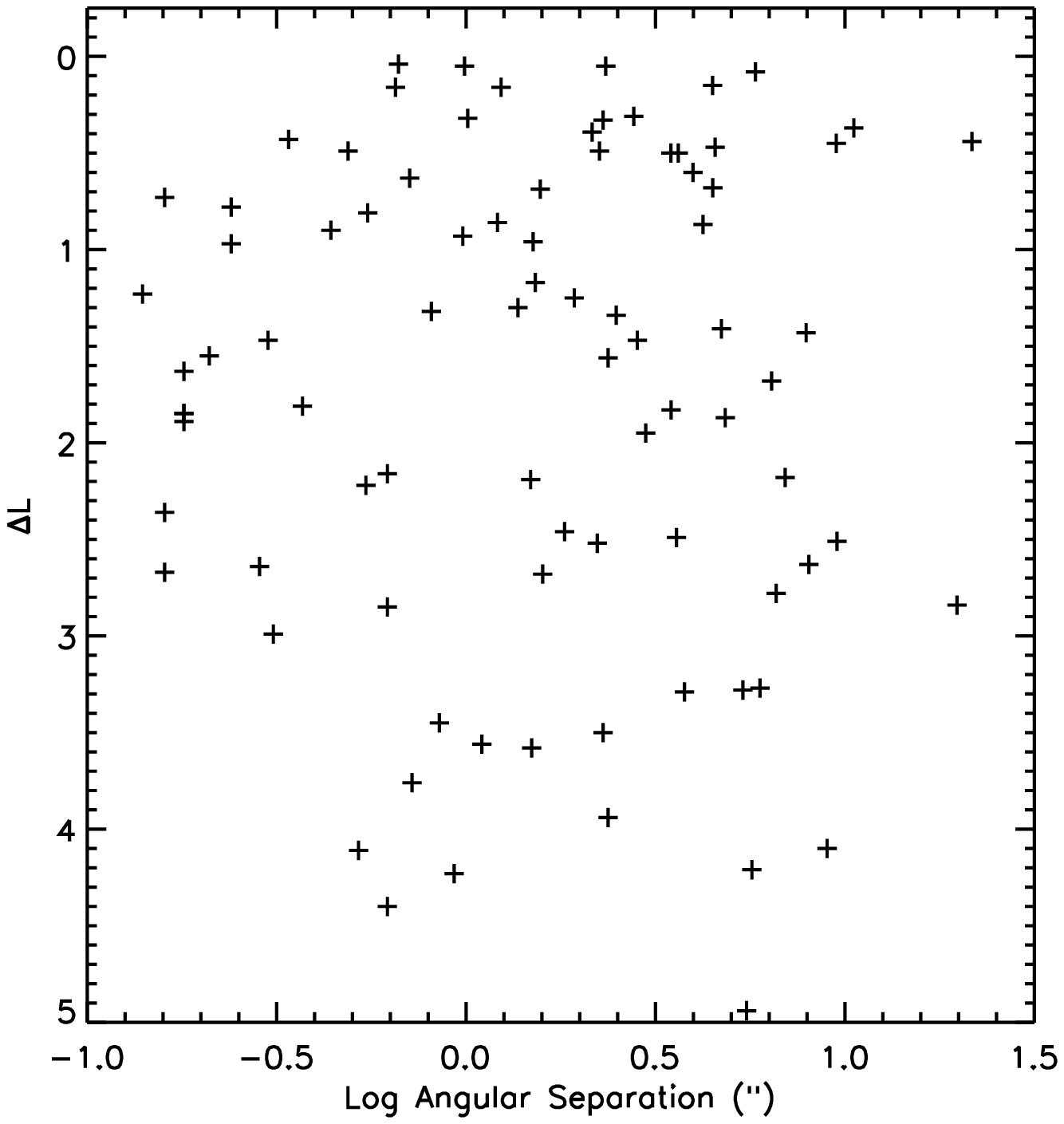}
\caption{Discovery Space.  These figures show the contrast of Class I binary companions versus angular separation (left) and versus log(angular separation/1\arcsec).  We only appear to be losing binary companions at a contrast higher than $\Delta$L$'=3$ and closer than 0\farcs5. \label{fig2.1}}
\end{figure}

\clearpage
\begin{figure}
\plotone{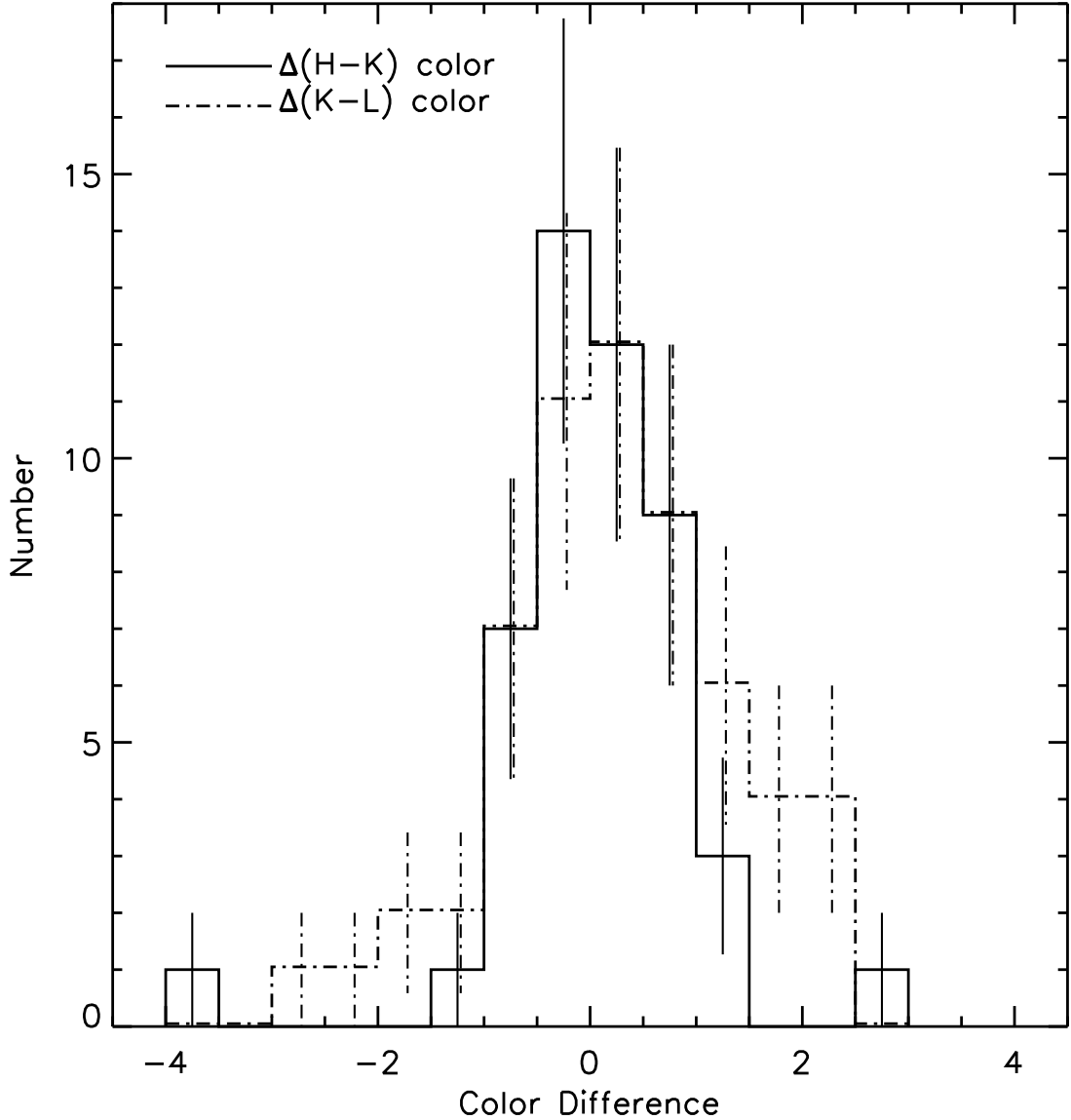}
\caption{The H-K and K-L$'$ color difference distributions. Both distributions are centered near a color difference of 0.  16\% of H-K color differences and 10\% of K-L$'$ color differences have color differences greater than the T Tauri system.  This percentage of Class I binaries with strong color differences is similar to the fraction of T Tauri stars with IR companions.  This figure does not include objects where we only have a lower limit on the H-band magnitude. \label{fig2.1}}
\end{figure}

\clearpage
\begin{deluxetable}{llrrllrrrcrcc}
\tabletypesize{\scriptsize}
\tablecaption{Source Characteristics}
\tablewidth{0pt}
\tablecolumns{10}
\tablehead{
\colhead{IRAS} &
\colhead{Associations} & 
\colhead{D(pc)\tablenotemark{a}} & 
\colhead{$L_{bol}$} & 
\colhead{$\alpha$(J2000)\tablenotemark{b}} &
\colhead{$\delta$(J2000)\tablenotemark{b}} & 
\colhead{J\tablenotemark{c}} & 
\colhead{H\tablenotemark{c}} & 
\colhead{K$_s$\tablenotemark{c}} & 
\colhead{$\alpha$\tablenotemark{d}} &

}
\startdata
00182+6223     & L1280                       & 4680(4)  &   366.9 & 00 20 56.79       &   +62 40 21.0  & 15.688 & 13.982 & 12.443 &  1.53  \\
00465+5028     & CB6, LBN 613 	             &  800(6)  &     8.8 & 00 49 24.50       &   +50 44 43.6  & 13.822 & 12.297 & 11.531 &  1.01  \\
00494+5617     & cluster in NGC 281          & 2940(21) &  9854.4 & 00 52 23.7$^{I}$  &   +56 33 45    & \nodata& \nodata& \nodata&  2.14  \\
01166+6635     &                             &  249(4)  &     0.4 & 01 20 03.93       &   +66 51 35.9  & 16.029 & 14.037 & 12.606 &  0.38  \\
02086+7600     & L1333                       &  180(5)  &     0.8 & 02 13 43.61       &   +76 15 06.0  & 13.715 & 12.254 & 11.193 &  0.88  \\
02232+6138     & cluster                     & 2040(46) &  74578. & 02 27 01.0$^{I}$  &   +61 52 14    & \nodata& \nodata& \nodata&  1.76  \\
02310+6133     & IC~1805                     & 2350(22) &  1029.9 & 02 34 48.79       &   +61 46 44.5  & \nodata& \nodata& 13.864 &  1.64  \\
02511+6023     & S190~HII                    & 2000(23) &   485.6 & 02 55 01.99       &   +60 35 41.7  & \nodata& 14.827 & 12.698 &  1.18  \\
\enddata
\end{deluxetable}


\clearpage
\begin{deluxetable}{lcclcc}
\tabletypesize{\scriptsize}
\tablecaption{Observing Nights}
\tablewidth{0pt}
\tablecolumns{4}
\tablehead{
\colhead{Date (UT)} &
\colhead{Observatory} & \colhead{Instr.} & 
\colhead{Weather} &  
}
\startdata

Nov. 2-5, 2003    &  UKIRT    &  UIST  & 4 nights photometric            \\
Nov. 29-30, 2003  &  UKIRT    &  UIST  & 1.5 nights lost                 \\
Feb. 11-15, 2004  &  UKIRT    &  UFTI  & 4 photometric half-nights       \\
May 15-17, 2004   &  UKIRT    &  UIST  & 3 half-nights lost              \\
May 24-25, 2004   &  UKIRT    &  UIST  & 2 half-nights, 0.25 night lost  \\
May 26-29, 2004   &  IRTF     &  SpeX  & 36 hours, photometric           \\
Jun. 19-21, 2004  &  UKIRT    &  UIST  & 0.5 night lost                  \\
Jul. 29, 2004     &  UH~2.2~m &  QUIRC & 1 night, photometric            \\
Aug. 2, 2004      &  Subaru   &  IRCS  & 0.5 nights lost, poor seeing    \\
Aug. 3-5, 2004    &  UH~2.2~m &  QUIRC & 3 nights, 1.75 nights lost      \\
Nov. 5, 2004      &  IRTF     &  SpeX  & 0.5 nights, clear               \\
Nov. 18, 2004     &  IRTF     &  SpeX  & 0.5 nights, clear               \\
Dec. 4-9, 2004    &  IRTF     &  SpeX  & 5 half-nights, 1.5 nights lost, poor seeing \\
Dec. 30-31, 2004  &  UKIRT    &  UIST  & 1.5 nights, 1 night lost        \\
Nov. 16-18, 2005  &  Subaru   &  CIAO  & 3 nights, 1 non-photometric     \\

\enddata

\end{deluxetable}

\clearpage

\clearpage
\begin{deluxetable}{lcccclccc}
\tabletypesize{\scriptsize}
\tablecaption{AO Observed Sample}
\tablewidth{0pt}
\tablecolumns{4}
\tablehead{
\colhead{IRAS} &
\colhead{r (\arcsec)\tablenotemark{a}} & \colhead{B\tablenotemark{b}} & 
\colhead{R\tablenotemark{b}} & \colhead{I\tablenotemark{b}} & \colhead{Date} &
}
\startdata

04240+2559   &   1.4  &   10.13  &   8.97  &    7.65  & 14Nov05 \\
04530+5126   &   0.5  &   17.80  &  13.69  &   11.35  & 14Nov05 \\
05289-0430E  &   0.4  &   17.51  &  14.96  &   14.54  & 15Nov05 \\
05289-0430W  &  17.1  &   17.51  &  14.96  &   14.54  & 15Nov05 \\
05302-0537   &  27.5  &   16.83  &  14.42  &   13.13  & 14Nov05 \\
05327-0457W  &  35.7  &   17.71  &  15.89  &   14.18  & 15Nov05 \\
05357-0650   &   0.3  &   12.98  &  10.52  &   10.38  & 15Nov05 \\
05375-0040   &   5.5  &   16.37  &  14.04  &   12.63  & 14Nov05 \\
\enddata
\tablenotetext{a}{The separation from the guide star to the target}
\tablenotetext{b}{The USNO magnitudes of the guide star}

\end{deluxetable}

\clearpage

\clearpage

\begin{deluxetable}{lcccccccccccc}
\tabletypesize{\scriptsize}
\tablecaption{Target Photometry}
\tablewidth{0pt}
\tablecolumns{7}
\tablehead{
\colhead{IRAS} & \colhead{$\#$\tablenotemark{a}} & 
\colhead{H} & \colhead{$\delta$H\tablenotemark{b}} & \colhead{date} &
\colhead{K} & \colhead{$\delta$K\tablenotemark{b}} & \colhead{date} &
\colhead{L$'$} & \colhead{$\delta$L$'$\tablenotemark{b}} & \colhead{date} &

}

\startdata

 00182+6223 & 1 &      14.71 &   0.14 &  28jul04 &  13.16 &   0.08 & 28jul04 &   11.36 &   0.05 &  01aug04 \\
 00182+6223 & 2 &      18.11 &   0.04 &  28jul04 &  15.39 &   0.05 & 28jul04 &   13.37 &   0.14 &  01aug04 \\
 00182+6223 & 3 &      15.11 &   0.05 &  28jul04 &  14.15 &   0.05 & 28jul04 &   13.01 &   0.08 &  01aug04 \\
 00465+5028 &   &      14.94 &   0.18 &  03nov03 &  12.65 &   0.08 & 04nov03 &    9.96 &   0.05 &  02nov03 \\
            &   &     \nodata& \nodata&  \nodata & \nodata& \nodata& \nodata &    9.79 &   0.04 &  19jun04 \\
 01166+6635 &   &      13.60 &   0.06 &  28jul04 &  12.30 &   0.06 & 28jul04 &   10.69 &   0.04 &  01aug04 \\
 02086+7600 &   &      11.90 &   0.04 &  28jul04 &  10.99 &   0.05 & 28jul04 & \nodata & \nodata&  \nodata \\  
 02310+6133 &   &     \nodata& \nodata&  \nodata & \nodata& \nodata& \nodata & \nodata & \nodata&  \nodata \\
\enddata

\tablenotetext{a}{The identifier of the L$'$ source in the finder chart}
\tablenotetext{b}{The photometric uncertaintly in this filter, as described in section 3.5}
\tablecomments{$^{*}$: Photometry includes flux from this source and an adjacent source, which were not resolved in this wavelength}

\end{deluxetable}

\clearpage

\clearpage
\begin{deluxetable}{lcccccccccccc}
\tabletypesize{\scriptsize}
\tablecaption{Binary Properties}
\tablewidth{0pt}
\tablecolumns{7}
\tablehead{
 \colhead{IRAS} &\colhead{$\#$\tablenotemark{a}}  & \colhead{d~(pc)} &\colhead{$\Delta$L$'$\tablenotemark{b}} & \colhead{r (\arcsec)\tablenotemark{c}} & \colhead{PA\tablenotemark{d}} & \colhead{Discoverer} 
}

\startdata

 03220+3035   &    &   290  &    1.30 &  1.37  & 200.3    &  Hodapp (1994) \\  
 F03258+3105  &    &   220  &    0.05 &  0.99  &  91.5    &  new \\  
 03260+3111   & 1  &   290  &    0.81 &  0.55  &  81.7    &  new \\  
 03260+3111   & 1  &   290  &    3.29 &  3.77  &  49.2    &  Haisch et al. (2004)\\ 
 03331+6256   & 1  &  1560  &    0.05 &  2.34  & 212.4    &  new \\                 
 04073+3800   & 1  &   350  &    5.18 & 12.94  &  28.6    &  Weintraub (1992)\\       
 04108+2803   & 2  &   140  &    0.44 & 21.64  &  64.5    &  Myers et al. (1987)\\    
 04113+2758   & 1  &   140  &    0.60 &  3.97  & 154.3    &  Kenyon et al. (1990) \\
\enddata

\tablenotetext{a}{The number of the primary star in the finder charts if there is more than one primary object per IRAS source.}
\tablenotetext{b}{The L$'$ magnitude difference between the primary and secondary stars.}
\tablenotetext{c}{The angular separation from the primary to the secondary star.}
\tablenotetext{d}{The position angle of the secondary star.}

\end{deluxetable}

\clearpage

\clearpage

\begin{deluxetable}{lcccccccccccc}
\tabletypesize{\scriptsize}
\tablecaption{Binary Detection Limits}
\tablewidth{0pt}
\tablecolumns{7}
\tablehead{
\colhead{} & \colhead{} & \colhead{} & \multicolumn{4}{c}{Inner Detection Limits (\arcsec)\tablenotemark{a}}  & \multicolumn{4}{c}{Outer Detection Limits (\arcsec)\tablenotemark{b}} \\
\colhead{IRAS} & \colhead{$\#$\tablenotemark{c}} &\colhead{d~(pc)} & 
\colhead{$\Delta$L$'=1$} & \colhead{$\Delta$L$'=2$} & \colhead{$\Delta$L$'=3$} & \colhead{$\Delta$L$'=4$} & \colhead{$\Delta$L$'=1$} & \colhead{$\Delta$L$'=2$} & \colhead{$\Delta$L$'=3$} & \colhead{$\Delta$L$'=4$} 
}

\startdata

00465+5028  &    &   800  &   0.25 & 0.30 & 0.72 & 999. &  6.25 &  6.25 &  6.25 &  6.25   \\
01166+6635  &    &   249  &   0.26 & 0.26 & 0.41 & 0.70 & 10.53 &  6.75 &  5.35 &  5.05   \\
02086+7600  &    &   180  &   0.39 & 999. & 999. & 999. & 27.78 & 20.19 & 16.01 & 15.13   \\
03220+3035  &    &   290  &   0.28 & 0.30 & 0.42 & 0.72 & 17.24 & 17.24 & 17.24 & 17.24   \\  
03225+3034  &    &   290  &   0.27 & 999. & 999. & 999. & 17.24 & 17.24 & 17.24 & 17.24   \\
F03258+3105 &    &   220  &   0.27 & 0.27 & 0.27 & 999. & 22.73 & 22.73 & 22.73 & 20.53   \\
03260+3111  & 1  &   290  &   0.30 & 0.33 & 0.72 & 0.96 & 17.24 & 17.24 & 17.24 & 17.24   \\
03260+3111  & 10 &   290  &   0.27 & 0.48 & 0.66 & 0.84 & 17.24 & 17.24 & 17.24 & 17.24   \\
\enddata

\tablenotetext{a}{The closest distance from the primary star that a fake companion star of the stated magnitude difference could be detected.  If inner detection limit is 999., then a companion of that contrast cannot be detected at any separation.}
\tablenotetext{b}{The farthest that a binary companion could be accepted.  This is limited by our 5000~AU separation limit or the 5\% contamination criterion.}
\tablenotetext{c}{The number of the primary star in the finder charts if there is more than one primary object per IRAS source.}

\end{deluxetable}


\clearpage

\begin{thebibliography}{}
\bibitem[()]{} 
\bibitem[Burnham (1978)]{Bur1978}Burnham, R., \emph{Burnham's Celestial Handbook}, Dover Publications, New York, 1978.

\bibitem[Chen \& Tokunaga (1994)]{Che1994} Chen, H. \& Tokunaga, A., 1994, \apjs, 90, 149

\bibitem[Cohen \& Kuhi (1979)]{Coh1979} Cohen, M. \& Kuhi, L., 1979, \apjs, 41, 743 

\bibitem[Connelley et al. (2007)]{Con2007} Connelley, M., Reipurth, B., \& Tokunaga, A., 2007, \aj, 133, 1528

\bibitem[Connelley et al. (2008)]{Con2008} Connelley, M., Reipurth, B., \& Tokunaga, A., 2008, \aj, in press

\bibitem[Correia et al. (2006)]{Cor2006} Correia, S., Zinnecker, H., Ratzka, T., \& Sterzik, M., 2006, \aap, 459, 909

\bibitem[Covey et al. (2006)]{Cov2006} Covey, K., Greene, T., Doppmann, G., Lada, C., AJ, 131, 512

\bibitem[Duch{\^e}ne et al. (2004)]{Duc2004} Duch{\^e}ne, G., Bouvier, J., Bontemps, S., Andr{\'e}, P., \& Motte, F., 2004, \aap, 427, 651

\bibitem[Duch{\^e}ne et al. (2007)]{Duc2006} Duch{\^e}ne, G., Delgado-Donate, E., Haisch, K., Loinard, L., \& Rodr\'\i guez, L., 2007, in Protostars and Planets V, eds B. Reipurth, D. Jewitt, \& K. Keil, University of Arizona Press, p.379

\bibitem[Duquennoy \& Mayor (1991)]{Duq1991} Duquennoy, A. \& Mayor, M., 1991, \aap, 248, 485

\bibitem[Ghez et al. (1991)]{Ghe1991} Ghez, A., Neugebauer, G., Gorham, P., Haniff, C., Kulkarni, S., Matthews, K., Koresko, C., \& Beckwith, S., 1991, \aj, 102, 2066

\bibitem[Haisch et al. (2004)]{Hai2004} Haisch, Jr., K., Greene, T., Barsony, M., \& Stahler, S., 2004, \aj, 127, 1747

\bibitem[Hartmann et al. (1999)]{Har1999} Hartmann, L.,  Calvet, N., Allen, L., Chen, H., Jayawardhana, R., 1999, \aj, 118, 1784 

\bibitem[Heintz (1969)]{Hei1969} Heintz, W., 1969, JRASC, 63, 275

\bibitem[Hodapp (1994)]{Hod1994}Hodapp, K., 1994, \apjs, 94, 615

\bibitem[Hodapp et al. (1996)]{Hod1996}Hodapp, K., Hora, J., Hall, D., et al. 1996, New Astronomy, 1, 177

\bibitem[Kenyon et al. (1990)]{Ken1990}Kenyon, S., Hartmann, L., Strom, K., \& Strom, S., 1990, \aj, 99, 869

\bibitem[Kobayashi et al. (2000)]{Kob2000}Kobayashi, N., Tokunaga, A., Terada, H., et al., 2000, Proc. SPIE 4008, eds. M. Iye \& A. Moorwood, 1056

\bibitem[Krisciunas et al. (1987)]{Kri1987}Krisciunas, K., 1987, \pasp, 99, 887

\bibitem[Lada (1991)]{Lad1991} Lada, C., 1991, in The Physics of Star Formation and Early Stellar Evolution, eds. C. J. Lada \& N. D. Kylafis, Kluwer Academic Publishers, p.329

\bibitem[Larson (2001)]{Lar2001} Larson, R., 2001, in IAU Symposium 200: The Formatio of Binary Stars, eds. H. Zinnecker \& R. Mathieu, Astronomical Society of the Pacific, p.93

\bibitem[Magnier et al. (1999)]{Mag1999} Magnier, E., Volp, A., Laan, K., van den Ancker, M., \& Waters, L., 1999, \aap, 352, 228

\bibitem[Mathieu et al. (2000)]{Mat2000a} Mathieu, R., Ghez, A., Jensen, E., \& Simon, M., 2000, in Protostars and Planets IV, eds. Mannings, Boss, \& Russell, University of Arizona Press, p.703

\bibitem[Mathis (2000)]{Mat2000b} Mathis, J., 2000, in  Allen's Astrophysical Quantities, ed. Cox, AIP Press

\bibitem[Meyer et al. (1997)]{Mey1997}  Meyer, M., Calvet, N., Hillenbrand, L., 1997, AJ, 114, 288

\bibitem[Mitchell (1767)]{Mit1767} Mitchell, J., 1767, Roy. Soc. Phil. Trans., 57, 234

\bibitem[Monet et al. (2003)]{Mon2003} Monet, D., Levine, S., Canzian, B., et al., 2003, \aj, 125, 984

\bibitem[Mundt et al. (1984)]{Mun1984} Mundt, R., B\"{u}ehrke, T., Fried, J., Neckel, T., Sarcander, M., Stocke, J., 1984, \aap, 140, 17 

\bibitem[Murakawa et al. (2004)]{Mur2004} Murakawa, K., Suto, H., Tamura, M., et al., 2004, \pasj, 56, 509 

\bibitem[Myers et al. (1987)]{Mye1987} Myers, P., Fuller, G., Mathieu, R., Beichman, C., Benson, P., Schild, R., \& Emerson, J., \apj, 319, 340

\bibitem[Osterloh et al. (1997)]{Ost1997} Osterloh, M., Henning, Th., \& Launhardt, R., 1997, \apjs, 110, 71

\bibitem[Patience et al. (2002)]{Pat2002} Patience, J., Ghez, A., Reid, I., \& Matthews, K., 2002, \aj, 123, 1570

\bibitem[Pich\'{e} et al. (1995)]{Pic1995} Pich\'{e}, F., Howard, E., Pipher, J., 1995, MNRAS, 275, 711 

\bibitem[Price et al. (2001)]{Pri2001} Price, S., Egan, M., Carey, S., Mizuno, D., \& Kuchar, T., 2001, \aj, 121, 2819

\bibitem[Ramsay Howat et al. (2004)]{Ram2004} Ramsay Howat, S., Todd, S., Leggett, S., et al., 2004, SPIE, 5492, 1160 

\bibitem[Rayner et al. (2003)]{Ray2003} Rayner, J., Toomey, D., Onaka, P., Denault, A., Stahlberger, W., Vacca, W., Cushing, M., Wang, S., 2003, \pasp, 115, 362

\bibitem[Reipurth et al. (2000)]{Rei2000} Reipurth, B., Yu, K., Heathcote, S., Bally, J., \& Rodr\'\i guez, L, 2000, \aj, 120, 1449 

\bibitem[Reipurth \& Zinnecker (1993)]{Rei1993} Reipurth, B. \& Zinnecker, H., 1993, \aap, 278, 81

\bibitem[Ressler \& Shure (1991)]{Res1991} Ressler, M. \& Shure, M., 1991, \aj, 102, 1398

\bibitem[Simons \& Tokunaga (2002)]{Sim2002} Simons, D. \& Tokunaga, A., 2002, PASP, 114, 169

\bibitem[Tokunaga \& Simons (2002)]{Tok2002} Tokunaga, A. \& Simons, D., 2002, PASP, 114, 180

\bibitem[Tokunaga et al. (1998)]{Tok1998} Tokunaga, A., Kobayashi, N., Bell, J., et al., 1998, SPIE, 3354, 512

\bibitem[Weintraub (1992)]{Wei1992} Weintraub, D., 1992, BAAS, 24, 1141

\bibitem[Yun \& Clemens (1994)]{Yun1994} Yun, J. \& Clemens, D., 1994, \apjs, 92, 145

\bibitem[Zinnecker \& Wilking (1992)]{Zin1992} Zinnecker, H., \& Wilking, B., 1992, in Binaries as Tracers of Stellar Formation, eds. A. Duquennoy \& M. Mayor, Cambridge University Press, p.269

\bibitem[()]{} 

\end{thebibliography}
\end{document}